\documentclass[conference,10pt]{IEEEtran}
\usepackage{epsfig,rotating,setspace,latexsym,amsmath,epsf,amssymb,amsfonts,bm,subfigure,epstopdf}
\usepackage{cite,authblk}
\usepackage{bbm}
\usepackage{mathtools,amsthm}
\usepackage{algorithm}
\usepackage[noend]{algpseudocode}
\usepackage{color}
\usepackage{systeme}

\algrenewcommand\algorithmicforall{\textbf{foreach}}
\algrenewcommand\algorithmicindent{.8em}

\DeclareMathOperator*{\argmax}{arg\,max}
\DeclareMathOperator*{\argmin}{arg\,min}

\newtheorem{theorem}{Theorem}
\newtheorem{lemma}{Lemma}

\IEEEoverridecommandlockouts
\allowdisplaybreaks
\begin{document}

\title{Age of Information of a Power Constrained Scheduler in the Presence of a Power \\ Constrained Adversary}

\author{Subhankar Banerjee \qquad Sennur Ulukus \qquad Anthony Ephremides\\
\normalsize Department of Electrical and Computer Engineering\\
\normalsize University of Maryland, College Park, MD 20742\\
\normalsize  \emph{sbanerje@umd.edu} \qquad \emph{ulukus@umd.edu} \qquad \emph{etony@umd.edu}}

\maketitle

\begin{abstract}
    We consider a time slotted communication network consisting of a base station (BS), an adversary, $N$ users and $N_{s}$ communication channels. In the first part of the paper, we consider the setting where $N_{s}$ communication channels $\mathcal{N}_{s}$ are heterogeneously divided among $N$ users. The BS transmits an update to the $i$th user on a subset of the communication channels $\mathcal{N}_{s,i}$  where $\mathcal{N}_{s,i}\cap \mathcal{N}_{s,j}$ is not necessarily an empty set. At each time slot, the BS transmits an update packet to a user through a communication channel and the adversary aims to block the update packet sent by the BS by blocking a communication channel. The BS has $n$ discrete transmission power levels to communicate with the users and the adversary has $m$ discrete blocking power levels to block the communication channels. The probability of successful transmission of an update packet depends on these power levels. The BS and the adversary have a transmission and blocking average power constraint, respectively. We provide a universal lower bound for the average age of information for this communication network. We prove that the uniform user choosing policy, the uniform communication channel choosing policy with any arbitrary feasible transmission power choosing policy is $4$ optimal; and the max-age user choosing policy, the uniform communication channel choosing policy with any arbitrary feasible transmission power choosing policy is $2$ optimal. In the second part of the paper, we consider the setting where the BS chooses a transmission policy and the adversary chooses a blocking policy from the set of randomized stationary policies and $\mathcal{N}_{s,i}=\mathcal{N}_{s}$ for all $i$, i.e., all users can receive updates on all channels. We show that a Nash equilibrium may or may not exist for this communication network, and identify special cases where a Nash equilibrium always exists.   
\end{abstract}

\section{Introduction}
We consider a wireless communication system consisting of $N$ users, one base station (BS), $N_{s}$ communication channels and an adversary. A communication channel can have different channel gains to different users, and thus, all the sub-carriers may not be available to all the users for transmission of an update packet. We consider the static setting. Thus, the communication channels are divided into $N$ potentially overlapping sets, where each set corresponds to a user. We denote the set of communication channels available to user $i$ as $\mathcal{N}_{s,i}$. A sub-carrier can be an element of multiple sets, and thus, the set $\mathcal{N}_{s,i}\cap \mathcal{N}_{s,j}$ is not necessarily empty. The cardinality of $\mathcal{N}_{s,i}$ is $N_{s,i}$. The set of all available channels is $\mathcal{N}_s= \bigcup_i\mathcal{N}_{s,i}$, and has cardinality $N_s$.  There are $n$ discrete power levels available to the BS for transmission of an update packet to the users and $m$ discrete power levels available to the adversary to block the transmission of an update packet. We consider a slotted time model. At each time slot, the BS chooses a transmission power to transmit an update packet to a user via a communication channel and the adversary chooses a communication channel and a blocking power to block any update packet that is being sent on the chosen channel.

A large amount of work has been done on the analysis of age of information for various applications and system models, such as, scheduling policies for wireless networks, gossip networks, caching systems, source coding problem, remote estimation, energy harvesting systems and many more, see e.g., 
\cite{kosta2017age, SunSurvey, YatesSurvey, kadota18, Najm17, Soysal19, Yates20_moments, Farazi18, Wu18, Leng19, Arafa19TwoHop, gu2020twohop, Arafa20, wireless-ephremides,  Buyukates19_hier, Elmagid20, Buyukates19_multihop, hsu18,  liu18,  Ceran18, Buyukates20_stragglers, yang20, ozfatura20, buyukates-fl, wang19_counting, bastopcu20_google, sun17_remote, chakravorty20, Bastopcu19_distortion, vaze21jsac,  Yates_Soljanin_source_coding, mayekar20, Bastopcu20_selective, Yates17sqrt, bastopcu2020LineNetwork, Eryilmaz21, Kaswan-isit2021, Yates21gossip, baturalp21comm_struc, bastopcu21gossip, kam20}. These papers consider systems without an adversary. The age of information in the presence of an adversary in a wireless communication network has been studied in the recent literature \cite{garnaev2019maintaining, nguyen2017impact,  xiao2018dynamic, banerjee2020fundamental, bhattacharjee2020competitive-short,banerjee2022age, banerjee2022game,kaswan2022age,kaswan2022susceptibility}. In particular, \cite{kaswan2022age,kaswan2022susceptibility} consider an adversarial gossip network. In this paper, we do not consider a gossip network, rather we consider that a central node, i.e., the BS transmits the update packets to the users. \cite{garnaev2019maintaining, nguyen2017impact} consider an adversary which decreases the signal to noise ratio of a communication link through jamming, due to which the rate of the communication decreases which results in a higher age for the communication system. In this paper, we consider that when the adversary blocks a communication channel it completely eliminates the update packet with a positive probability. \cite{xiao2018dynamic} considers an adversary which blocks the communication channel for a duration in time which increases the average age of the system by disabling communication in that interval. In this paper, we consider that the adversary blocks the communication channel in a time slotted manner. \cite{banerjee2020fundamental, bhattacharjee2020competitive-short} consider an adversary which completely eliminates the update packet, however, they do not consider any power constraint on the adversary. In this paper, we consider a power constrained adversary. \cite{banerjee2022age, banerjee2022game} consider a power constrained adversary which completely eliminates the update packet. They have considered that on the time horizon $T$, the adversary blocks $\alpha T$ time slots where $0<\alpha <1$. On the contrary, in this paper, we consider that at each time slot $t$, the adversary chooses one of the $m$ blocking power levels with a pmf $\bm{d}(t)$ and the expected power to be less than or equal to a power constraint. Different than the adversary in \cite{banerjee2022age, banerjee2022game}, the adversary in this paper completely eliminates the update packet with a positive probability (strictly less than $1$), and this probability depends on the blocking power chosen by the adversary and the transmission power chosen by the BS.

In the first part of this paper, we propose algorithms to minimize the average age of information for the described wireless communication network. We show that the uniform user choosing policy together with the uniform communication channel choosing policy and any arbitrary feasible transmission power choosing policy is $4$ optimal, and in a special case, it is $2$ optimal. We show that the maximum-age user choosing policy together with the uniform communication channel choosing policy and any arbitrary feasible transmission power choosing policy is $2$ optimal. 

In the second part of this paper, we relax the system model and consider that at each time slot the BS can choose any one of the $N_{s}$ sub-carriers for transmission of an update packet to any one of the $N$ users, i.e., $\mathcal{N}_{s,i}=\mathcal{N}_{s}$, for all $i$. We also restrict the action space of the BS and the action space of the adversary only to the stationary policies. If the power level choosing algorithms are not fixed for the BS and for the adversary and if those are included in the action space of the BS and the action space of the adversary, then we show that in the stationary policy regime a Nash equilibrium may not exist. We give a counter example to prove this. We also show a special case in which the Nash equilibrium exists. However, when the power level choosing algorithms for the BS and for the adversary are fixed, i.e., those are not included in the list of the actions of the BS and the list of the actions of the adversary, then the Nash equilibrium always exists. 

\section{System Model and Problem Formulation}
At each time slot, the BS schedules a user $i$ out of $N$ users, $N>1$, with a user choosing algorithm $\pi_{u}$ and chooses a communication channel out of $N_{s,i}$ communication channels, $N_{s,i}>1$, with a communication channel choosing algorithm $\pi_{s}$ to transmit an update packet to the scheduled user $i$. In this paper, we use sub-carrier and communication channel interchangeably. We consider that $n$ discrete transmission powers, namely $\{p_{1},p_{2},\cdots,p_{n}\}$ are available to the BS, and at each time slot the BS chooses one of these $n$ transmission powers, following a power choosing algorithm $\pi_{p}$. Thus, an action of the BS is a triplet $(\pi_{u}, \pi_{s}, \pi_{p})$ and we call a valid triplet as a BS scheduling algorithm $\bm{\pi}$. We call the set of all causal scheduling algorithms as ${\bm{\Pi}}$. Let us consider that $\pi_{p}$ is such that at time slot $t$ the BS chooses the $i$th transmission power with probability $e_{i}(t)$. We consider the following power constraint for the BS,
\begin{align}\label{eq:bspow}
    \sum_{i=1}^{n} e_{i}(t) p_{i} \leq \bar{p}, \quad {t\in{\{1,\cdots,T\}}}
\end{align}

We consider that an adversary is present in the system as well. At each time slot, the adversary chooses a sub-carrier out of $N_{s}$ sub-carriers following an algorithm $\psi_{s}$ to block any update packet that is being transmitted by the BS in that sub-carrier. We consider that $m$ discrete blocking powers, namely $\{p_{1}', p_{2}', \cdots,p_{m}'\}$ are available to the adversary and at each time slot the adversary chooses one of these powers, following a blocking power choosing algorithm $\psi_{p}$, to block any update packet on the sub-carrier chosen by $\psi_{s}$. Thus, an action of the adversary is a pair $(\psi_{s}, \psi_{p})$ and we call a valid pair as an adversarial action $\bm{\psi}$. We call the set of all valid adversarial actions as $\bm{\Psi}$. Let us consider that $\psi_{p}$ is such that at time slot $t$, the adversary chooses the $i$th blocking power with probability $d_{i}(t)$. We consider the following power constraint for the adversary,
\begin{align}\label{eq:advpow}
    \sum_{i=1}^{m} d_{i}(t) p'_{i} \leq \tilde{p}, \quad {t\in{\{1,\cdots,T\}}}
\end{align}

We create an $n \times m$ matrix $\bm{F}$, whose $(i,j)$th element, $f_{i,j}$, represents the probability of successful transmission of an update packet corresponding to the BS transmission power $p_{i}$ and adversary blocking power $p_{j}'$. Thus, at time slot $t$ if the BS schedules the user $k$, and chooses the sub-carrier $l$ to transmit an update packet with power $p_{i}$ and if the adversary blocks the sub-carrier $l$ with power $p_{j}'$, then with probability $f_{i,j}$ the age of the $k$th user at time slot $(t+1)$ becomes $1$ and with probability $1-f_{i,j}$ the age of the $k$th user at time slot $t+1$ increases by one. 

The age of user $i$ at time slot $t$ is defined as $t-t_{i}(t)$, where $t_{i}(t)$ is the last time slot when the $i$th user has successfully received an update packet. Note that the minimum value for the age of user $i$ is $1$. We consider that at each time slot the BS has a fresh update packet to transmit for every user present in the system. Here by fresh update packet, we mean the update packet for the $i$th user at time slot $t$ is generated at time slot $t$. As we are interested in freshness, we assume that if the $i$th user does not receive the corresponding update packet at time slot $t$, then that update packet gets dropped at the BS without any cost. This is a valid assumption used in  \cite{banerjee2020fundamental, bhattacharjee2020competitive-short, banerjee2022game, banerjee2022age}.

The adversary has the knowledge of $\pi_{u}$, $\pi_{s}$ and $\pi_{p}$. However, as the BS uses a randomized algorithm at time slot $t$, the adversary has no knowledge about which user will get scheduled, which sub-carrier will get chosen and which transmission power will get used at time slot $t'$ when $t\leq t'\leq T$. However, at time slot $t$ it has full knowledge about all these for time slot $t'$ when $1\leq t'< t$, and the adversary can optimize its future actions based on these available information. The adversary has full knowledge about the elements of each set $\mathcal{N}_{s,i}$. The age of user $i$ at time slot $t$ corresponding to a BS scheduling algorithm $\bm{\pi}$ and adversarial action $\bm{\psi}$ is denoted as $v_{i}^{(\bm{\pi}, \bm{\psi})}(t)$, thus, $v_{i}^{(\bm{\pi}, \bm{\psi})}(t) =  t- t_{i}(t)$, and the expected age of user $i$ at time slot $t$, is denoted as $\Delta_{i}^{(\bm{\pi}, \bm{\psi})}(t)$. Note that, if the BS successfully transmits an update packet to user $i$ at time slot $t$, then $v_{i}^{(\bm{\pi}, \bm{\psi})}(t+1)=1$, otherwise $v_{i}^{(\bm{\pi}, \bm{\psi})}(t+1)=v_{i}^{(\bm{\pi}, \bm{\psi})}(t)+1$. The average age of the overall system corresponding to the BS scheduling algorithm $\bm{\pi}$ and adversarial action $\bm{\psi}$ is,
\begin{align}\label{eq:3:obj}
    \Delta^{(\bm{\pi}, \bm{\psi})} = \limsup_{T \to \infty} \frac{1}{T} \sum_{t=1}^{T} \frac{1}{N} \sum_{i=1}^{N} \Delta_{i}^{(\bm{\pi}, \bm{\psi})}(t)
\end{align}

For the simplicity of presentation, in the rest of the paper we ignore the superscript $(\bm{\pi},\bm{\psi})$, unless we specify otherwise. Now, as the BS has no control over the adversary, we consider the following constrained optimization problem,
\begin{align}\label{eq:mainobj}
    \Delta^{*} = \sup_{\bm{\psi}\in{\bm{\Psi}}} \inf_{\bm{\pi}\in\bm{{\Pi}}}  & \quad \Delta^{(\bm{\pi},\bm{\psi)}} \nonumber \\  
     \textrm{s.t.} &  \quad (\ref{eq:bspow}), (\ref{eq:advpow})
\end{align}

For the second part of the paper, we consider a relaxed system model. We consider that at each time slot, all the $N_{s}$ sub-carriers are available to the BS to transmit an update packet to any one of the $N$ users, i.e., $\mathcal{N}_{s,i}=\mathcal{N}_{s}$ for all $i$. The BS chooses a scheduling algorithm and the adversary chooses an adversarial action from the corresponding sets of stationary randomized policies. In other words, $\pi_{u}$ is such that at each time slot the BS chooses a user following a pmf $\bm{u} = [u_{1}, u_{2}, \cdots, u_{N}]$, $\pi_{s}$ is such that at each time slot the BS chooses a sub-carrier following a pmf $\bm{s} = [s_{1}, s_{2}, \cdots, s_{N_{s}}]$ and $\pi_{p}$ is such that at each time slot the the BS chooses a power following a pmf $\bm{e} = [e_{1}, e_{2}, \cdots, e_{n}]$. Similarly, $\psi_{s}$ is such that at each time slot the adversary blocks a sub-carrier following a pmf $\bm{a} = [a_{1}, a_{2}, \cdots, a_{N_{s}}]$ and $\psi_{p}$ is such that at each time slot the adversary chooses a blocking power following a pmf $\bm{d} = [d_{1}, d_{2}, \cdots, d_{m}]$. Thus, the power constraints for the adversary and the BS become $\sum_{i=1}^{m} d_{i} p'(i) \leq \tilde{p}$ and $\sum_{i=1}^{n} e_{i} p(i) \leq \bar{p}$, respectively. When we restrict ourselves only to the stationary randomized policies, instead of writing $\Delta^{\bm{\pi},\bm{\psi}}$ as in (\ref{eq:3:obj}), we write the average age of the overall system corresponding to pmfs $\bm{u}$, $\bm{s}$, $\bm{e}$ (these three pmfs are chosen by the BS) and the pmfs $\bm{a}$, $\bm{d}$ (these two pmfs are chosen by the adversary) as $\Delta^{\bm{u}, \bm{s}, \bm{e}, \bm{a}, \bm{d}}$. We denote the expected age of user $i$ at time slot $t$ as $\Delta_{i}^{\bm{u}, \bm{s}, \bm{e}, \bm{a}, \bm{d}} (t)$. Thus, the average age for the $i$th user becomes
\begin{align}
    \Delta^{\bm{u}, \bm{s}, \bm{e}, \bm{a}, \bm{d}}_{i} = \limsup_{T\to\infty} \frac{1}{T} \sum_{t=1}^{T} \Delta_{i}^{\bm{u}, \bm{s}, \bm{e}, \bm{a}, \bm{d}} (t)
\end{align}

Let us assume that the set of all valid user choosing pmfs, the set of all valid sub-carrier choosing pmfs and the set of all valid transmission power choosing pmfs are $\mathcal{F}_{u}$, $\mathcal{F}_{s}$ and $\mathcal{F}_{e}$, respectively. Similarly, the set of all valid sub-carrier blocking pmfs and the set for all valid blocking power choosing pmfs are $\mathcal{F}_{a}$ and $\mathcal{F}_{d}$, respectively. For a given adversarial action, namely a sub-carrier blocking pmf $\bm{a}$, and a blocking power level choosing pmf $\bm{d}$, the BS aims to minimize the average age of the overall system by selecting a scheduling algorithm, namely a user choosing pmf $\bm{u}$, a sub-carrier choosing pmf $\bm{s}$ and a transmission power choosing pmf $\bm{e}$ from the set $B(\bm{a},\bm{d})$, where $B(\bm{a},\bm{d})$ is defined as follows,
\begin{align}
    B(\bm{a}, \bm{d}) = \argmin_{(\bm{u}\in{\mathcal{F}_{u}}, \bm{s}\in{\mathcal{F}_{s}}, \bm{e}\in{\mathcal{F}_{e}}, \sum_{i=1}^{n} e_{i} p_{i} \leq \bar{p})} \Delta^{\bm{u}, \bm{s}, \bm{e}, \bm{a}, \bm{d}}
\end{align}
Similarly, for a given scheduling algorithm, i.e., a triplet of pmfs $(\bm{u}, \bm{s}, \bm{e})$, the adversary aims to maximize the average age by choosing a pair of pmfs, namely $(\bm{a}, \bm{d})$ from the set $B(\bm{u}, \bm{s}, \bm{e})$, where $B(\bm{u}, \bm{s}, \bm{e})$ is defined as
\begin{align}
    B(\bm{u}, \bm{s}, \bm{e}) = \argmax_{(\bm{a}\in{\mathcal{F}_{a}},          \bm{d}\in{\mathcal{F}_{d}},   \sum_{i=1}^{m} d_{i} p'(i) \leq \tilde{p})} \Delta^{\bm{u}, \bm{s}, \bm{e}, \bm{a}, \bm{d}} 
\end{align}

We call a 5-tuple of pmfs, namely $({\bm{u}}, {\bm{s}}, {\bm{e}}, {\bm{a}}, {\bm{d}})$ as a Nash equilibrium point if and only if $({\bm{u}},{\bm{s}}, {\bm{e}})\in {B({\bm{a}}, {\bm{d}})}$ and $({\bm{a}}, {\bm{d}})\in{B({\bm{u}},{\bm{s}}, {\bm{e}})}$. 

In the previous Nash equilibrium setting we consider that the transmission power choosing pmf $\bm{e}$ and blocking power  choosing pmf $\bm{d}$ are components of the action space of the BS and the action space of the adversary, respectively. However, if $\bm{e}$ and $\bm{d}$ are fixed and not included in the action space of the BS and the action space of the adversary, respectively, then we define,
\begin{align}
    B(\bm{a}) = \argmin_{(\bm{u}\in{\mathcal{F}_{u}}, \bm{s}\in{\mathcal{F}_{s}})} \Delta^{\bm{u}, \bm{s}, \bm{e}, \bm{a}, \bm{d}}
\end{align}
Similarly, we write,
\begin{align}
    B(\bm{u}, \bm{s}) = \argmax_{(\bm{a}\in{\mathcal{F}_{a}})} \Delta^{\bm{u}, \bm{s}, \bm{e}, \bm{a}, \bm{d}} 
\end{align}
We call a triplet of pmfs, namely $({\bm{u}}, {\bm{s}}, {\bm{a}})$ as a Nash equilibrium point if and only if $({\bm{u}},{\bm{s}})\in {B({\bm{a}}, )}$ and ${\bm{a}}\in{B({\bm{u}},{\bm{s}})}$.

\section{Algorithm and Analysis of Age}
We find a fundamental lower bound for the optimization problem in (\ref{eq:mainobj}). Let us define $x= \argmax_{i\in{\{1,\cdots,m\}}} p'_{i} \leq \tilde{p} $. Consider the following adversarial action: at each time slot the adversary blocks any one of the $N_{s}$ sub-carriers with a uniform pmf and chooses the power level $p_{x}$. We denote this adversarial action as $\bar{\bm{\psi}}=(\bar{\psi}_{s}, \bar{\psi}_{p})$. At each time slot, if the BS schedules the user which has the maximum age and breaks the tie with scheduling the lower indexed user, we call that user choosing policy as the max-age policy. (In this paper, we will present our results in a sequence of lemmas and theorems, with some explanations. The proofs are skipped here due to space limitations, and will be provided in the journal version.)

\begin{lemma}\label{lemma:max_age}
    For the adversarial action $\bar{\bm{\psi}}$, an optimal user choosing policy is the max-age policy; and if the $i$th user gets chosen by the max-age policy, then an optimal sub-carrier choosing policy is to choose a sub-carrier in $\mathcal{N}_{s,i}$ uniformly.
\end{lemma}

Let us define $\bar{y} = \argmin_{i\in{\{1,\cdots,n\}}} p_{i} \geq \bar{p}$.

\begin{theorem}\label{first_th}
The average age of the communication network defined in (\ref{eq:3:obj}) is lower bounded by $\frac{(N+1) N_{s}}{2(N_{s}-1 + f_{\bar{y},x})}$.
\end{theorem}

Now, we consider that at each time slot the BS schedules a user $i$ with probability $\frac{1}{N}$ and chooses one of the $N_{s,i}$ sub-carriers with probability $\frac{1}{N_{s,i}}$, to transmit an update packet to the scheduled user with transmission power $p_{y}$ with probability $\beta$ and with transmission power $p_{\bar{y}}$ with probability $(1-\beta)$, where $\beta$ satisfies the following identity:
\begin{align}\label{eq:betapower}
    \beta p_{y} + (1- \beta) p_{\bar{y}} = \bar{p}
\end{align}
Let us denote this BS scheduling policy as $\hat{\tilde{\bm{\pi}}}$. Let us define $\bar{x} = \argmin_{i\in{\{1,\cdots,m\}}} p_{i}' \geq \tilde{p}$.

\begin{theorem}\label{sec_th}
The average age of the communication system when the BS employs the scheduling algorithm $\hat{\tilde{\bm{\pi}}}$ is upper bounded by $2 N$; when $N_{s,i}=N_{s}$ for all $i$, then the average age is upper bounded by $\frac{N N_{s}} {N_{s} - 1 + \beta f_{y,\bar{x}} + (1-\beta) f_{\bar{y}, \bar{x}}} $.
\end{theorem}

Now, we consider that at each time slot the BS schedules the max-age user, $i$, and chooses one of the $N_{s,i}$ sub-carriers with probability $\frac{1}{N_{s,i}}$. We also consider that the BS chooses power $p_{y}$ with probability $\beta$ and power $p_{\bar{y}}$ with probability $1-\beta$, where $\beta$ satisfies (\ref{eq:betapower}). Denote this BS scheduling policy as $\tilde{\tilde{\bm{\pi}}}$.

\begin{theorem}\label{th:max_age_upper}
The average age of the communication system when the BS employs the scheduling algorithm $\tilde{\tilde{\bm{\pi}}}$ is upper bounded by $\frac{(N+1) \bar{N}_{s}}{2 (\bar{N}_{s} - 1 + \beta f_{y,\bar{x}} + (1-\beta) f_{\bar{y},\bar{x}})}$, where $\bar{N}_{s} = \min {\{N_{s,1}, N_{s,2}, \cdots, N_{s,N}\}}$.
\end{theorem}

Next, we make some concluding remarks about the findings of this section. From Theorem~\ref{first_th} and Theorem~\ref{sec_th}, we see that in the general setting, $\hat{\tilde{\bm{\pi}}}$ is $\frac{4 N  (N_{s} - 1 + f_{\bar{y},x})}{(N+1) N_{s}}$ optimal, where
\begin{align}
    \frac{4 N  (N_{s} - 1 + f_{\bar{y},x})}{(N+1) N_{s}} \leq 4
\end{align}

For the special case, when $\mathcal{N}_{s,i}=\mathcal{N}_{s}$, for all $i$, $\hat{\tilde{\bm{\pi}}}$ is $\frac{2 (N+1) (N_{s}-1 + f_{\bar{y},x})}{N(N_{s}-1 + f_{y,\bar{x}})}$ optimal, where
\begin{align}
    \frac{2 (N+1) (N_{s}-1 + f_{\bar{y},x})}{N(N_{s}-1 + f_{y,\bar{x}})} \leq & \frac{2 (N_{s}-1 + f_{\bar{y},x})}{(N_{s}-1 + f_{y,\bar{x}})} \\ 
    \leq & \frac{2 N_{s}}{N_{s}-1} \label{eq:2opt} \\ 
    \leq& 4
\end{align}
If $N_{s}$ is large, then the right side of (\ref{eq:2opt}) can be approximated as $2$. Thus, for the aforementioned special case and for large $N_{s}$, $\hat{\tilde{\bm{\pi}}}$ is $2$ optimal.

From Theorem~\ref{first_th} and Theorem~\ref{th:max_age_upper}, we  see that the scheduling policy $\tilde{\tilde{\bm{\pi}}}$ is $\frac{\bar{N}_{s}} {\bar{N}_{s}-1}$ optimal and as ${N}_{s,i}>1$, for all $i$,  $\tilde{\tilde{\bm{\pi}}}$ is $2$ optimal.  Note that when $\bar{p}$ exactly matches with one of the powers from the sets $\{p_{1},p_{2},\cdots,p_{n}\}$ and $\mathcal{N}_{s,i}=\mathcal{N}_{s}$, for all $i$, then $\tilde{\tilde{\bm{\pi}}}$ is the optimal scheduling policy.

\section{Equilibrium Points of the Average Age for Randomized Stationary Action Space}
Let us assume that at each time slot the BS chooses a user following a pmf $\bm{u}$, chooses a sub-carrier following a pmf $\bm{s}$, chooses a transmission power with a pmf $\bm{e}$ and the adversary chooses a sub-carrier with a pmf $\bm{a}$ and chooses a blocking power following a pmf $\bm{d}$. Recall that for this section we use a relaxed system model, where we consider that $\mathcal{N}_{s,i}=\mathcal{N}_{s}$, for all $i$. At some time slot $t$, user $i$ successfully receives an update packet transmitted by the BS and then after waiting for $\Gamma_{i}$ time slots it again receives another update packet from the BS. Note that $\Gamma_{i}$ is a random variable. The evolution of the age for the $i$th user is a renewal process and $\Gamma_{i}$ is a renewal interval. Thus, from the renewal reward theorem,
\begin{align}\label{eq:ren}
   \Delta_{i}^{\bm{u}, \bm{s}, \bm{e}, \bm{a}, \bm{d}} = \frac{\mathbb{E}\left[\Gamma_{i}^{2} + \Gamma_{i}\right]}{2 \mathbb{E}\left[\Gamma_{i}\right]} 
\end{align}
Let the probability of successful transmission of the update packet to user $i$ be $q_{i}$. Then, $\Gamma_{i}$ is geometrically distributed with success probability $q_{i}$. Thus, (\ref{eq:ren}) simplifies as,
\begin{align}
     \Delta_{i}^{\bm{u}, \bm{s}, \bm{e}, \bm{a}, \bm{d}} = \frac{1}{q_{i}}
\end{align}
\begin{theorem}\label{lemma:1}
The optimal sub-carrier choosing pmf $\bm{s}$, for a given adversarial action, namely, a pair of pmfs $(\bm{a}, \bm{d})$, depends only on $\bm{a}$ and is independent of user choosing pmf $\bm{u}$, transmission power choosing pmf $\bm{e}$ and $\bm{d}$. Moreover, if the adversary blocks any $l$ sub-carriers with lowest probability then the optimal choice for the BS is to choose any subset of these $l$ sub-carriers with probability $1$. Similarly, the optimal user scheduling pmf $\bm{u}$ does not depend on $\bm{a}$, $\bm{s}$, $\bm{d}$, $\bm{e}$. The optimal user scheduling pmf is the uniform pmf. 
\end{theorem}

\begin{theorem}\label{lemma:2}
The optimal sub-carrier blocking pmf, $\bm{a}$, for a given BS scheduling policy depends only on $\bm{s}$ and is independent of $\bm{u}$, $\bm{e}$ and $\bm{d}$. Moreover, if the BS chooses any $l$ sub-carriers with the highest probability, then the optimal choice for the adversary is to block any subset of these $l$ sub-carriers with probability $1$.
\end{theorem}

Without loss of generality, let $p_{1}\leq p_{2}\leq\cdots\leq p_{n}$ and $p'_{1}\leq p'_{2}\leq\cdots\leq p'_{m}$. Thus, we have $f_{1,j}\leq f_{2,j}\leq \cdots \leq f_{n,j}$ and $f_{i,1}\geq f_{i,2}\geq\cdots\geq f_{i,m}$, $i=1,\cdots,n$, $j=1,\cdots,m$. Algorithm~\ref{alg:cap} below provides an optimal transmission power choosing pmf $\bm{e}$ for a given blocking power choosing pmf $\bm{d}$. The algorithm states that, if $\bar{p} < p_{1}$, then there does not exist a feasible $\bm{e}$; if $p_{n}< \bar{p}$, then the optimal $\bm{e}$ is to choose the power $p_{n}$ with probability $1$;  If these two cases do not occur, then we define  $x = \argmax_{i\in{\{1,\cdots,n}\},p_{i}<\bar{p}}i$ and $y=\argmin_{i\in{\{1,\cdots,n\}},p_{i}> \bar{p}}i$. Clearly, $x<y$. We define a constant, $g_{i} = \sum_{j=1}^{m} d_{j} f_{i,j}$, $i=1,\cdots,n$. We call the constant $\left(g_{i} + g_{x} \frac{p_{y}-p_{i}} {p_{x} - p_{y}} - g_{y} \frac{p_{x}-p_{i}}{p_{x} - p_{y}}\right)$ as the coefficient for power $p_{i}$, $i\in{\{1,\cdots,n\}}\backslash\{x,y\}$. Then, we traverse from power $p_{y+1}$ to power $p_{n}$, we call this procedure as the first traversing procedure. During this traversing process, if we find that $\left(g_{j} + g_{x} \frac{p_{y}-p_{j}} {p_{x} - p_{y}} - g_{y} \frac{p_{x}-p_{j}}{p_{x} - p_{y}}\right)$, $j>y$, is a strictly positive number, then we change the coefficient of the power $p_{k}$ as  $\left(g_{k} + g_{x} \frac{p_{j}-p_{k}} {p_{x} - p_{j}} - g_{j} \frac{p_{x}-p_{k}}{p_{x} - p_{j}}\right)$, $k\in{\{1,\cdots,n\}}\backslash\{x,j\}$. We keep on doing this procedure till we reach $p_{n}$. Let us assume that during this traversing procedure $p_{i}$ is the last power for which we get a positive coefficient, then we define $y=i$. Then, we start performing a second traversing procedure from the power $p_{x-1}$ to the power $p_{1}$. During this traversing process, if we find that the coefficient of $p_{l}$, $l<x$, is a strictly positive number, then we change the coefficient of the power $p_{k}$ as  $\left(g_{k} + g_{l} \frac{p_{y}-p_{k}} {p_{l} - p_{y}} - g_{y} \frac{p_{l}-p_{k}}{p_{l} - p_{y}}\right)$, $k\in{\{1,\cdots,n\}}\backslash\{l,y\}$. We keep on doing this procedure till we reach $p_{1}$. Let us assume that during this second traversing procedure $p_{r}$ is the last power for which we get a positive coefficient, then we define $x=r$. Now, if $\bar{p}$ exactly matches one of the powers from the set $\{p_{1},p_{2},\cdots,p_{n}\}$, without loss of generality assume that $p_{i} = \bar{p}$, then we compare the two vectors $\bm{z}_{i}$ and $\left(\frac{\bar{p} - p_{y}}{p_{x}-p_{y}} \bm{z}_{x} + \frac{p_{x}-\bar{p}}{p_{x}-p_{y}} \bm{z}_{y}\right)$ and select the one which minimizes (\ref{eq:ren}), otherwise we select $\left(\frac{\bar{p} - p_{y}}{p_{x}-p_{y}} \bm{z}_{x} + \frac{p_{x}-\bar{p}}{p_{x}-p_{y}} \bm{z}_{y}\right)$, where $\bm{z}_{i}$ is the $i$th basis vector of $\mathbb{R}^{n}$. 

We note that, Algorithm~\ref{alg:cap} finds an optimal solution in $O(n)$ time. Next, we state the optimality of Algorithm~\ref{alg:cap}.

\begin{algorithm}[t]
\caption{For a given $\bm{d}$ finding an optimal $\bm{e}$}\label{alg:cap}
\begin{algorithmic}
\State \textbf{Inputs}: $\bm{d}$, $\bm{F}$, $\bm{p}$, $\bar{p}$
\State \textbf{Define}: $\bm{g}=(g_{1},g_{2},\cdots,g_{n})$, where $g_{i} = \sum_{j=1}^{m} d_{j} f_{i,j}$, $x = \argmax_{i\in{\{1,2,\cdots,n}\},p_{i}<\bar{p}}i$ and $y=\argmin_{i\in{\{1,2,\cdots,n\}},p_{i}> \bar{p}}i$, $\bm{z}_{i}$ is the $i$th basis vector for $\mathbb{R}^{n}$, $x_{1}=x$, $y_{1}=y$
\If{$\bar{p}<p_{1}$} 
    \State \textbf{Return}: Solution does not exist 
    
    \ElsIf{$p_{n}<\bar{p}$}
        \State \textbf{Return}: $\bm{z}_{n}$ 
    \EndIf
    
    \For{$i= y+1:n $}
        \If{$\left(g_{i} + g_{x} \frac{p_{y} - p_{i}}{p_{x}-p_{y}} - g_{y} \frac{p_{x}-p_{i}}{p_{x}-p_{y}} \right)> 0$}
        \State $y=i$
        
        \EndIf
    \EndFor
    \For{$i=1:x-1$}
         \If{$\left(g_{i} + g_{x} \frac{p_{y} - p_{i}}{p_{x}-p_{y}} - g_{y} \frac{p_{x}-p_{i}}{p_{x}-p_{y}} \right)> 0$}
        \State $x=i$
        
        \EndIf
    \EndFor
    \State \textbf{Define}: $\bm{e} = \left(\frac{\bar{p} - p_{y}}{p_{x}-p_{y}} \bm{z}_{x} + \frac{p_{x}-\bar{p}}{p_{x}-p_{y}} \bm{z}_{y}\right)$
    \If{$x_{1}+1=y_{1}-1$}
        \If{$\sum_{i=1}^{n} e_{i} \sum_{j=1}^{m} d_{j} f_{i,j} \leq \sum_{j=1}^{m} d_{j} f_{x_{1}+1,j}$}
            \State \textbf{Return}: $\bm{z}_{x+1}$
        \Else
            \State \textbf{Return}: $\bm{e}$
            
        \EndIf
    \Else
        \State \textbf{Return}: $\bm{e}$
    \EndIf   
\end{algorithmic}
\end{algorithm}

\begin{theorem}\label{th:alg:cap}
 For a given blocking power pmf $\bm{d}$, Algorithm~\ref{alg:cap} gives an optimal transmission power pmf $\bm{e}$. 
\end{theorem}

\begin{algorithm}[t]
\caption{For a given $\bm{e}$ finding an optimal $\bm{d}$}\label{alg:2}
\begin{algorithmic}
\State \textbf{Inputs}: $\bm{e}$, $\bm{F}$, $\bm{p}$, $\bar{p}$
\State \textbf{Define}: $\bm{g}=(g_{1},g_{2},\cdots,g_{m})$, where $g_{i} = \sum_{j=1}^{n} e_{j} f_{j,i}$, $x = \argmax_{i\in{\{1,2,\cdots,m}\},p_{i}'<\tilde{p}}i$ and $y=\argmin_{i\in{\{1,2,\cdots,m\}},p_{i}'>\tilde{p}}i$, $\bm{z}_{i}$ is the $i$th basis function for $\mathbb{R}^{n}$, $x_{1}=x$, $y_{1}=y$
\If{$\tilde{p}<p_{1}'$} 
    \State \textbf{Return}: Solution does not exist 
    
    \ElsIf{$p_{n}'<\tilde{p}$}
        \State \textbf{Return}: $z_{n}$ 
    \EndIf

    \For{$i= y+1:n $}
        \If{$\left(g_{i} + g_{x} \frac{p_{y}' - p_{i}'}{p_{x}' -p_{y}'} - g_{y} \frac{p_{x}'-p_{i}'}{p_{x}'-p_{y}'} \right)< 0$}
        \State $y=i$
        
        \EndIf
    \EndFor
    \For{$i=1:x-1$}
         \If{$\left(g_{i} + g_{x} \frac{p_{y}' - p_{i}'}{p_{x}'-p_{y}'} - g_{y} \frac{p_{x}'-p_{i}'}{p_{x}'-p_{y}'} \right)< 0$}
        \State $x=i$
        
        \EndIf
    \EndFor
   \State \textbf{Define}: $\bm{d} = \left(\frac{\tilde{p} - p_{y}'}{p_{x}' -p_{y}'} \bm{z}_{x} + \frac{p_{x}'-\tilde{p}}{p_{x}' -p_{y}' } \bm{z}_{y}\right)$
   \If{$x_{1}+1 = y_{1}-1$}
        \If{$\sum_{j=1}^{m} d_{j} \sum_{i=1}^{n} e_{i} f_{i,j} \leq \sum_{i=1}^{n} e_{i} f_{i,x_{1}+1}$}
            \State \textbf{Return}: $\bm{d}$
        \Else
            \State \textbf{Return}: $\bm{z}_{x+1}$
            
        \EndIf
        
    \Else   
        \State \textbf{Return}: $\bm{d}$
    \EndIf
\end{algorithmic}
\end{algorithm}

Algorithm~\ref{alg:2} provides an optimal blocking power choosing pmf $\bm{d}$ for a given $\bm{e}$. In Algorithm~\ref{alg:2}, we perform a similar traversing procedure as Algorithm~\ref{alg:cap}. The only difference is while traversing in Algorithm~\ref{alg:cap}, we change the coefficient of a power level if the corresponding coefficient is strictly positive, in Algorithm~\ref{alg:2}, we change the coefficient if it is strictly negative. Next, we state the optimality of Algorithm~\ref{alg:2}.

\begin{theorem} \label{thm7}
For a given transmission power choosing pmf $\bm{e}$, Algorithm~\ref{alg:2} gives an optimal blocking power pmf $\bm{d}$. 
\end{theorem}

Next, we present a counter example which suggests that when the transmission power choosing pmf and the blocking power choosing pmf are not fixed and are part of the action space of the BS and the action space of the adversary, respectively, then a Nash equilibrium may not exist. Consider a system where the BS has three power levels and the adversary has also three power levels, i.e., $n=m=3$. Both the power constraint for the BS and the adversary is $3.5$ watts. The feasible powers for the BS and for the adversary are the same, which is $[1, 3, 5]$. The matrix $\bm{F}$ is chosen as
\begin{align} 
    \bm{F}=\begin{bmatrix}
 0.5 & 0.35 & 0.2\\0.6 & 0.55 & 0.4\\0.8 & 0.7 & 0.65
\end{bmatrix}
\end{align}
We can show that for this example, for a given $\bm{d}$, $\bm{e}$ cannot be of the form $[e_{1}, e_{2}, e_{3}]$, where $e_{i}>0$, ${i}\in{\{1,2,3\}}$ and satisfy $\sum_{i=1}^{3} e_{i} p_{i} \leq \bar{p}$. Now, from Algorithm~\ref{alg:cap}, we know that if the adversary chooses powers $3$ and $5$, then the optimal choice for the BS is to choose powers $3$ and $5$, similarly, if the adversary chooses powers $1$ and $5$, then the optimal choice for the BS is to choose powers $1$ and $5$. From Algorithm~\ref{alg:2}, we know that if the BS chooses powers $1$ and $5$, then the optimal choice for the adversary is to choose powers $3$ and $5$, similarly, if the BS chooses powers $3$ and $5$, then the optimal choice for the adversary is to choose powers $1$ and $5$. Thus, a Nash equilibrium does not exist for this example.

In the next theorem, we consider the Nash equilibrium when the transmission power choosing pmf and the blocking power choosing pmf are not included in the action space of the BS and in the action space of the adversary, respectively.

\begin{theorem}\label{th:nash}
The triplet of actions $(\hat{\bm{u}},\hat{\bm{s}} ,\hat{\bm{a}})$ is the Nash equilibrium point, where $\hat{\bm{a}}$ and $\hat{\bm{s}}$ are the uniform pmfs over $N_{s}$ sub-carriers and $\hat{\bm{u}}$ is the uniform pmf over $N$ users. 
\end{theorem}

Next, we present a special case in which the Nash equilibrium exists even when the transmission power choosing pmf and the blocking power choosing pmf are part of the action space of the BS and the action space of the adversary, respectively. Consider that the matrix $\bm{F}$ has the property,
\begin{align}
 f_{i,j} - f_{1,j} = l_{i}, \quad {j\in{\{1,\cdots,m\}}}, ~ {i\in{\{1,\cdots,n\}}}
\end{align}
where $l_{i}$ are non-negative constants. Consider a fixed blocking power choosing pmf ${\bm{d}}$. Then, $g_{i}$ in Algorithm~\ref{alg:cap} is
\begin{align}
    g_{i} = \sum_{j=1}^{m} {d}_{j} f_{i,j} 
    =  \sum_{j=1}^{m} {d}_{j}f_{1,j} +l_{i}
\end{align}
Thus,
\begin{align}
    g_{i} &+ g_{x}\frac{p_{y}-p_{i}}{p_{x}-p_{y}} - g_{y} \frac{p_{x}-p_{i}}{p_{x}-p_{y}} \nonumber\\ 
    = & \left(\sum_{j=1}^{m} {d}_{j}f_{1,j}\right)\left(1 + \frac{p_{y}-p_{i}}{p_{x}-p_{y}}- \frac{p_{x}-p_{i}}{p_{x}-p_{y}}\right)  + l_{x}  \frac{p_{y}-p_{i}}{p_{x}-p_{y}}  \nonumber \\ & - l_{y} \frac{p_{x}-p_{i}}{p_{x}-p_{y}} + l_{i}
\end{align}
Thus, the sign of $g_{i} + g_{x}\frac{p_{y}-p_{i}}{p_{x}-p_{y}} - g_{y} \frac{p_{x}-p_{i}}{p_{x}-p_{y}}$ does not depend on $\bm{{d}}$, which implies that the optimal transmission power choosing pmf is the same for all $\bm{d}$. Similarly, the sign of $g_{i} + g_{x} \frac{p_{y}' - p_{i}'}{p_{x}'-p_{y}'} - g_{y} \frac{p_{x}' - p_{i}'}{p_{x}'-p_{y}'}$ in Algorithm~\ref{alg:2} does not depend on $\bm{e}$, in other words the optimal blocking power choosing pmf is independent of $\bm{e}$. Now, run Algorithm~\ref{alg:cap} for any arbitrary $\bm{d}$ and denote the output as $\hat{\bm{e}}$, similarly run Algorithm~\ref{alg:2} for any arbitrary $\bm{e}$ and denote the output as $\hat{\bm{d}}$. Then, using Theorem~\ref{th:nash}, we have that the $5$-tuple $(\hat{\bm{b}}, \hat{\bm{c}}, \hat{\bm{e}}, \hat{\bm{a}}, \hat{\bm{d}})$ is the unique Nash equilibrium.

\bibliographystyle{unsrt} 
\bibliography{references-infocom2023}
\end{document}